\documentclass[
]{ceurart}

\usepackage{listings}
\begin{document}


\copyrightyear{2025}
\copyrightclause{Copyright for this paper by its authors.
  Use permitted under Creative Commons License Attribution 4.0
  International (CC BY 4.0).}

% \conference{Workshop on Meta-Research in HCI, CHI'25, April 26, 2025, Yokohama, Japan}
\conference{Meta-HCI '25: First Workshop on Meta-Research in HCI, April 26, 2025, Yokohama, Japan}

\title{UX Remix: Improving Measurement Item Design Process Using Large Language Models and Prior Literature}

% \tnotetext[1]{You can use this document as the template for preparing your
  % publication. We recommend using the latest version of the ceurart style.}

\author[1]{Hyeonggeun Yun}[%
% orcid=0000-0002-0877-7063,
email=hg.russ.yun@companoid.io,
% url=https://yamadharma.github.io/,
]
% \cormark[1]
\address[1]{Companoid Labs, Seoul, Republic of Korea}

\author[1]{Jinkyu Jang}[%
% orcid=0000-0002-0877-7063,
email=jk.chairman@companoid.io,
% url=https://yamadharma.github.io/,
]
\cormark[1]
% \address[1]{Companoid Labs, Seoul, Republic of Korea}

\cortext[1]{Corresponding author.}
% \fntext[1]{These authors contributed equally.}

\begin{abstract}
  Researchers often struggle to develop measurement items and lack a standardized process. To support the design process, we present UX Remix, a system to help researchers develop constructs and measurement items using large language models (LLMs). UX Remix leverages a database of constructs and associated measurement items from previous papers. Based on the data, UX Remix recommends constructs relevant to the research context. The researchers then select appropriate constructs based on the recommendations. Afterward, selected constructs are used to generate a custom construct, and UX Remix recommends measurement items. UX Remix streamlines the process of selecting constructs, developing measurement items, and adapting them to research contexts, addressing challenges in the selection and reuse of measurement items. This paper describes the implementation of the system, the potential benefits, and future directions to improve the rigor and efficiency of measurement design in human-computer interaction (HCI) research.
\end{abstract}

\begin{keywords}
    User experience\sep
    Constructs\sep
    Measurement items\sep
    Evaluation\sep
    Large language models
\end{keywords}

\maketitle

\section{Introduction}

As the scope of computing has expanded, evaluating user experience (UX) in interactive systems has become increasingly important. One of the most widely adopted methods for evaluating UX is conducting  studies with interactive systems and subsequently collect survey data regarding constructs from users \cite{perrig2024measurement}. Therefore, to design a survey for the study, researchers must create measurement items for the specific constructs they want to evaluate. To develop measurement items, they often leverage measurement items used in prior research \cite{ perrig2024measurement, darin2019instrument, lallemand2017could, zheng2022ux, muller2014survey}. The simplest approach is to utilize well-established questionnaires that are commonly used in human-computer interaction (HCI) research \cite{darin2019instrument, lallemand2017could, muller2014survey}. For instance, researchers can utilize a system usability scale (SUS) \cite{bangor2008empirical}, a NASA task load index (NASA-TLX) \cite{hart2006nasa}, a user burden scale (UBS) \cite{suh2016developing}, a questionnaire for user interaction satisfaction (QUIS) \cite{chin1988development}, and a user experience questionnaire (UEQ) \cite{laugwitz2008construction} to measure the well-known constructs such as usability, usefulness, and ease-of-use. Another approach is to adopt measurement items of constructs from previous studies in related fields \cite{perrig2024measurement, law2014attitudes}. However, finding and adopting similar constructs and measurement items from related research is challenging \cite{perrig2024measurement, graser2024identifying, schrepp2020comparison}. In addition, when the research context differs from previous research, existing measurement items may not be entirely appropriate. In this case, researchers should modify existing measurement items or develop new items \cite{perrig2024measurement, muller2014survey, graser2024identifying, bargas2011old, hornbaek2006current}.

Designing measurement items lacks a standardized process, so researchers should explore various studies to collect and modify measurement items \cite{perrig2024measurement, law2014attitudes, hornbaek2006current, lindgaard2013introduction}. In some cases, measurement items are not publicly available and there are often no guidelines to modify measurement items \cite{perrig2024measurement}. Consequently, researchers should manually investigate relevant studies that have addressed similar constructs, extract measurement items, and adapt them to their specific research context \cite{perrig2024measurement, muller2014survey, graser2024identifying, bargas2011old, hornbaek2006current}. This process is both time-consuming and resource-intensive \cite{vermeeren2010user}. Furthermore, instead of reusing the measurement items developed in previous research, researchers often develop new measurement items \cite{perrig2024measurement, lindgaard2013introduction}. However, this practice can compromise the reliability and rigor of the measurement items. Inefficiencies and inconsistencies inherent in the measurement item design process highlight the need for methodological improvements in HCI research.

Despite the challenges associated with the design of measurement items in HCI research, there has been limited exploration of systems to assist this process \cite{zheng2024evalignux}. Therefore, we introduce a system, UX Remix, supporting the design of measurement items for evaluation in HCI research. UX Remix uses constructs and associated measurement items from prior research. UX Remix then generates a custom construct and measurement items tailored to the evaluation purpose through an LLM. Researchers can utilize the system to customize the construct aligned with their research contexts and develop measurement items accordingly. We expect researchers to be able to improve the process of finding relevant constructs and developing measurement items for evaluation using UX Remix.

In this paper, we introduce the detailed measurement item design process with UX Remix. First, we describe the process of collecting previous literature and extracting information to build a database of constructs and their measurement items. Subsequently, we present the implementation of UX Remix and its core pages using the collected items. Finally, we discuss the potential and future directions of UX Remix.

\section{Related Works}

Surveys are widely used to assess the effectiveness of interactive computing systems \cite{perrig2024measurement, darin2019instrument, muller2014survey, syahrozad2024evaluation}. To evaluate UX, selecting the appropriate measurement items for surveys is an important yet challenging \cite{graser2024identifying, schrepp2020comparison, hillman2023understanding, hodrien2021review}. Consequently, evaluation has often relied on long-established questionnaires \cite{perrig2024measurement, darin2019instrument, muller2014survey, hillman2023understanding} such as SUS \cite{bangor2008empirical}, NASA-TLX \cite{hart2006nasa}, UEQ \cite{laugwitz2008construction}, QUIS \cite{chin1988development}, UBS \cite{suh2016developing}, and more. Although these surveys are credible and widely adopted, they primarily focus on general constructs such as usability, utility, and ease of use \cite{zheng2022ux, zheng2024evalignux}. As a result, they are often insufficient for measuring more diverse aspects of UX \cite{darin2019instrument, lallemand2017could, graser2024identifying, vermeeren2010user, mortazavi2024exploring}. Therefore, previous research has explored and proposed methods for designing surveys and measurement items specific to domains \cite{graser2024identifying, hillman2023understanding} such as mixed reality \cite{alexandrovsky2021evaluating, tcha2016questionnaire}, robotics \cite{ye2024autonomy, putten2018development}, conversational agents \cite{zheng2022ux, yun2020chatbot, rese2024perceived}, and smart home \cite{borgert2023home, schenkluhn2024connecting}.

Designing constructs and measurement items that align with specific research contexts still remains a difficult work, because researchers adopt different approaches to designing measurement items, including modifying existing items or developing new ones \cite{perrig2024measurement, darin2019instrument, muller2014survey, graser2024identifying}. Perrig et al. analyzed 85 distinct questionnaires in 153 papers from proceedings of the ACM Conference on Human Factors in Computing Systems (ACM CHI conference proceedings) published between 2019 and 2022 \cite{perrig2024measurement}. Most of these questionnaires were used only once (70.59\%), and 55\% of the papers modified their measurement items to fit their research by rewriting items, deleting irrelevant items or utilizing only items appropriate to the research context. Darin et al. emphasized that in survey-based evaluation, researchers should carefully select measurement items that align with evaluation objectives and apply them in an appropriate context \cite{darin2019instrument}. They highlighted the importance of considering the reuse of existing validated measurement items and adopting approaches that extend or adapt them as needed. Previous studies indicate that researchers frequently encounter difficulties in finding suitable measurement items, often resorting to compiling items from various studies or developing new ones \cite{perrig2024measurement, darin2019instrument, muller2014survey, graser2024identifying}. These processes are time-consuming and resource-intensive \cite{vermeeren2010user}, underscoring the need for more standardized and accessible measurement practices.

To address the challenges of designing measurement items in evaluation, recent work by Zheng et al. proposed a system that supports metric exploration for evaluation planning \cite{zheng2024evalignux}. Their system leveraged large language models (LLMs) to help researchers explore appropriate metrics for their evaluation plans. The system enabled researchers to design comprehensive and systematic evaluation plans. Their research demonstrated the potential of LLMs to recommend and design suitable metrics for evaluation. However, their system primarily focused on metric exploration and did not address the development of measurement items.

Based on the literature review, our research focuses on the process for designing constructs and measurement items. We introduce a system that utilizes an LLM to recommend constructs and develop measurement items tailored to specific evaluation purposes. The researchers can then design a custom construct and develop its measurement items using the system. We aim to enhance the re-usability of measurement items from previous studies and facilitate the development of measurement items for evaluation.

\section{UX Remix}

\begin{figure}
  \centering
  \includegraphics[width=\textwidth]{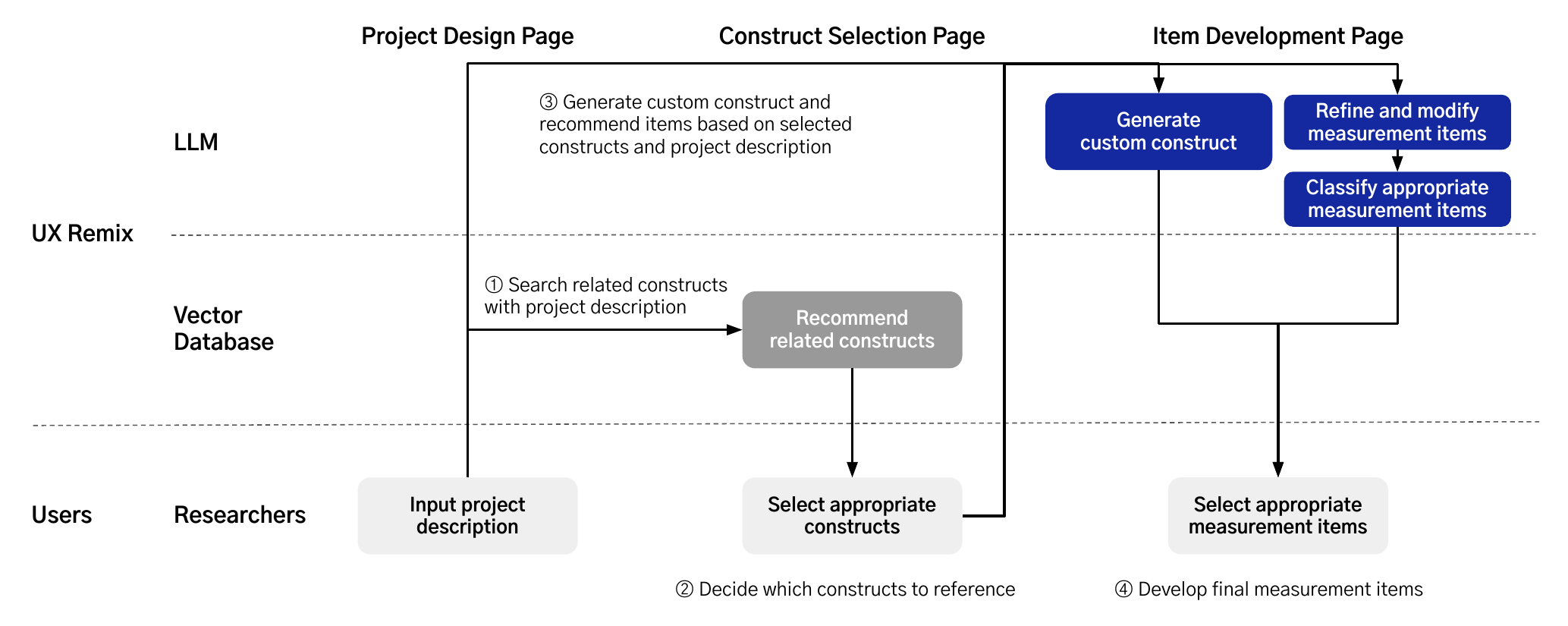}
  \caption{In UX Remix, users (researchers) fill out the project descriptions. (1) The vector database searches for 10 related constructs and recommends them to users based on the project description. (2) Users select the appropriate constructs that would be used as prompts in a large language model (LLM). (3) LLM generates a custom construct and measurement items for evaluation. (4) Users develop measurement items using recommendations of the LLM.}
  \label{fig:teaser}
\end{figure}

% UX evaluation을 위한 UX metric과 measurement items의 선택과 설계를 지원하기 위해 우리는 UX Remix을 개발했습니다. UX Remix는 이전의 연구에서 시행된 survey기반의 UX metric과 measurement items을 기반으로, 현재 연구 목적에 어울리는 metric과 its measurement items을 추천합니다. 이러한 추천을 바탕으로 researchers는 자신의 연구 목적에 부합하는 UX evalaution metric을 커스텀화하여 설계할 수 있고, measurement item을 구성할 수 있습니다. 

% 이를 위해 우리는 먼저, 이전의 HCI 연구 논문으로부터 survey 기반의 UX evalaution metric과 measurement items을 수집했습니다. 총 185개의 논문으로부터 697개의 UX evaluation metric과 그에 대한 measurement items이 발견되었습니다. 우리는 ChatGPT의 GPTs 기능을 활용하여 각각의 논문에서 UX metric과 measurement items, 그리고 논문에 언급된 metric definition, metric의 usage 등을 수집하였습니다. 수집한 정보는 Vector Database에 저장되었고, 이러한 데이터를 기반으로 UX Remix가 구현되었습니다. UX Remix는 연구자로부터 시스템의 정보, UX 평가 목적, 시스템의 핵심 요소, UX evalaution 요소를 입력받고, 이러한 정보를 embedding하고, vector search를 통해 가장 유사한 UX metric 10개를 추천합니다. 추천한 10개의 metric에 대해 researchers는 definition, 이전 연구의 usage, likert type and scale, measurement items list를 확인할 수 있습니다. 그 중 연구자가 자신의 evaluation purpose에 어울리는 metric을 선택할 수 있고, 그 선택을 토대로 시스템은 LLM을 활용해 custom metric의 이름과 definition을 작성합니다. 그리고, 선택한 metric의 measurement items를 현재 연구 context에 맞게 수정한 후, 그 중 evaluation purpose와 일치하거나 일치하지 않는 item을 분류하여 연구자에게 추천합니다. 연구자는 이를 기반으로 자신의 연구에 어울리는 measurement item을 구성할 수 있습니다. The details are below. 

To support the design process of the constructs and measurement items for evaluation, we developed UX Remix. Based on constructs and measurement items from the previous literature, UX Remix recommends appropriate constructs and measurement items that align with current research objectives. Using these recommendations, researchers can customize constructs and configure measurement items to suit their specific research context.

To develop the system, we first collected survey-based constructs and measurement items from previous HCI literature. We utilized an LLM to extract information about the constructs and pre-process them. The collected constructs and measurement items were stored in a vector database, which served as a foundation for implementing UX Remix. UX Remix receives input from researchers about the project description, and the input is used for a vector search to recommend the 10 most relevant constructs. The researchers can then select the constructs that best align with their evaluation purposes. Based on the selected constructs, the system utilizes an LLM to generate a custom construct, refine the measurement items, and classify them as aligned or misaligned with the purpose of the evaluation, providing the researchers with recommendations accordingly. The researchers can then configure the most appropriate measurement items for their research. The details of UX Remix are below.

% To achieve this, we first collected survey-based constructs and measurement items from previous HCI research papers. A total of 697 constructs and their corresponding measurement items were identified from 185 research papers. We utilized an LLM to extract information about the constructs and pre-process them. The collected constructs and measurement items were stored in a vector database, which served as a foundation for implementing UX Remix.

% UX Remix operates by receiving input from researchers about the project description. These inputs are embedded and processed through a vector search to recommend the 10 most relevant constructs. For each of the recommended constructs, researchers can review the specific information, such as a definition, a previous usage, a scale type, a range, and a list of measurement items. Researchers can then select the constructs that best align with their evaluation purposes.

% Based on the selected constructs, the system utilizes an LLM to generate a custom construct. Furthermore, it modifies the measurement items of the selected constructs to fit the current research context. After that, the system classifies the measurement items as aligned or misaligned with the purpose of the evaluation, providing researchers with recommendations accordingly. Researchers can then configure the most appropriate measurement items for their research. The details of UX Remix are below.

\subsection{Data Collection and Preprocessing}

To collect constructs and measurement items, we gathered papers published in HCI-related conferences and journals: (1) ACM SIGCHI-sponsored conferences and journals, (2) SCIE (Science Citation Index Expanded) journals such as International Journal of Human-Computer Studies, Interacting with Computers, Computers in Human Behavior, etc. (3) Other conference and journal papers in the fields of information science, cognitive science, and psychology. Therefore, we used the following search query to find literature:

\texttt{("questionnaire" OR "scale" OR "measurement items" OR "items" OR  "metric" OR "measurement") AND "Likert" AND (HCI OR Human-Computer Interaction) AND "[journal or conference name]"}

The query was used to find various constructs and measurement items in HCI-related papers. It consists of keywords on measures (\textit{questionnaire, scale, measurement items, items, metric, measurement, Likert}), field (\textit{HCI, Human-Computer Interaction}), and name of publications. The papers were searched in Google Scholar and the ACM Digital Library. Each paper was reviewed by the authors. If constructs and measurement items were explicitly mentioned, they were included in the paper database. Even if the same construct name or measurement items appeared in different papers, they were collected separately. As a result, a total of 697 constructs were identified from 185 research papers.

% To collect constructs and measurement items, we gathered papers published in HCI-related conferences and journals (e.g., ACM SIGCHI-sponsored conferences and journals, IEEE conferences and journals, International Journal of Human-Computer Studies, Computers in Human Behavior, and others). These papers were searched in Google Scholar and the ACM Digital Library using various combinations of the keywords such as ``HCI'' + ``Questionnaire,'' or ``Scale,'' or ``Measurement Items,'' or ``Items'' + ``[journal or conference name].''
% Each paper was reviewed by the authors, and if constructs and measurement items were explicitly mentioned and provided, they were included in the paper database. Even if the same construct name or measurement items appeared in different papers, they were collected separately. As a result, a total of 697 constructs were identified from 185 papers.

To extract information from the papers, we utilized a ChatGPT GPT-4o model. Based on the previous paper using an LLM to analyze metrics \cite{zheng2024evalignux}, we designed the GPT to extract the constructs and information of the papers. The GPT extracted the following constructs and paper-related information: construct name, construct definition, construct usage, point and type of measurement items, measurement items, the number of measurement items, title of the paper, and APA format of the paper. We extracted the information from each paper and stored them in the paper database. The details of our GPT are described in the Appendix~\ref{appendix:gpt_design}.

After extracting information from the papers, we pre-processed the measurement items of the data. Since the measurement items were originally designed in the context of previous studies, we revised them by replacing specific contextual terms with the generalized term ``[Evaluation Target]''. We also utilized an LLM to perform data pre-processing. The rewording of the measurement items was carried out using the following prompt: ``Please rewrite the above items by replacing the specific term like [system name or feature] with a generalized term like [Evaluation Target]. If there is no specific term, do not change the items. Keep the structure and format of the sentences unchanged''. A total of 697 constructs' measurement items were reworded accordingly. For instance, the item ``I could get a robot to perform a specific task.'' was modified to ``I could get a [Evaluation Target] to perform a specific task.''

To utilize the collected data for construct recommendation, the constructs were stored in a vector database. We used the Qdrant database as the vector database, storing all the extracted constructs and information from the papers. In the vector database, embedding vectors were generated using details of the constructs such as construct name, construct definition, construct usage, and measurement items to facilitate vector-based similarity search. To generate embedding vectors, we utilized a ``text-embedding-004'' model of Google. The embedding vectors have a dimension of 768, and cosine similarity was employed for vector search.

\subsection{System Design and Implementation}

% 우리의 UX Remix System은 웹 기반의 시스템으로 구현되었다. 이 시스템의 목표는 researchers나 practitioners가 UX metric과 measurement items을 쉽게 설계할 수 있도록 돕는 것이다. 따라서, 우리는 총 3가지의 section을 통해서 researchers과 practitioners가 UX metric을 선택하고, Measurement items를 구성할 수 있게 시스템을 개발하였다. 3가지의 섹션은 다음과 같다.: (1) Project Design Section, (2) Metric Selection Section, (3) Item Construction Section
Our UX Remix system was implemented as a web-based system. The primary objective of the system is to help researchers easily design constructs and measurement items in evaluation. To achieve the purpose, we developed the system with three core pages, enabling researchers to select the constructs and develop the measurement elements effectively (see Figure~\ref{fig:teaser}). The three pages are as follows: (1) project design page, (2) construct selection page, and (3) item development page.

\subsubsection{Project Design Page}

% Project Design Page에서 researchers와 practitioners는 프로젝트에 대한 기본 정보를 입력합니다. 먼저 Project Title을 입력하고, 그 후에 System Information, UX Evaluation Purpose, Interactive features of the system과 UX Evaluation Construct를 입력합니다. System Information은 평가하려는 System 또는 Product가 가장 잘 설명하는 category와 type에 기반한 정보를 입력합니다. (e.g. AI-powered Emotional Chatbot) UX Evaluation Purpose에서는 research를 통해 탐구하고 싶은 key aspects를 입력합니다. (e.g. I want to study how the anthropomorphism of an AI chatbot affects users' trust.) Evaluation Purpose 입력이 끝나면, 이 정보를 기반으로 Evaluation Design 정보를 입력합니다. Evaluation Design 정보는 Evaluation Purpose를 기반으로 Interactive Features와 UX Evaluation Construct를 입력할 수 있습니다. 이는 각각 hypothesis에서 Independent variable과 dependent variable의 역할을 하게 됩니다. 앞선 예시처럼 "I want to study how the anthropomorphism of an AI chatbot affects users' trust"를 Evaluation Purpose로 입력했다면, Interactive Features는 Anthropomorphism이 될 수 있고, UX Evaluation Construct는 Trust가 될 수 있습니다. 이처럼, evaluation purpose를 기반으로 각각을 입력하게 됩니다. 모든 정보의 입력이 끝나면,  시스템은 그들이 입력한 정보로 기본적인 프로젝트의 기본 정보를 저장합니다. 이 정보는 향후 metric 선택과 measurement item 구성을 위한 추천에서 활용합니다.

\begin{figure}[h]
  \centering
  \includegraphics[width=0.95\linewidth]{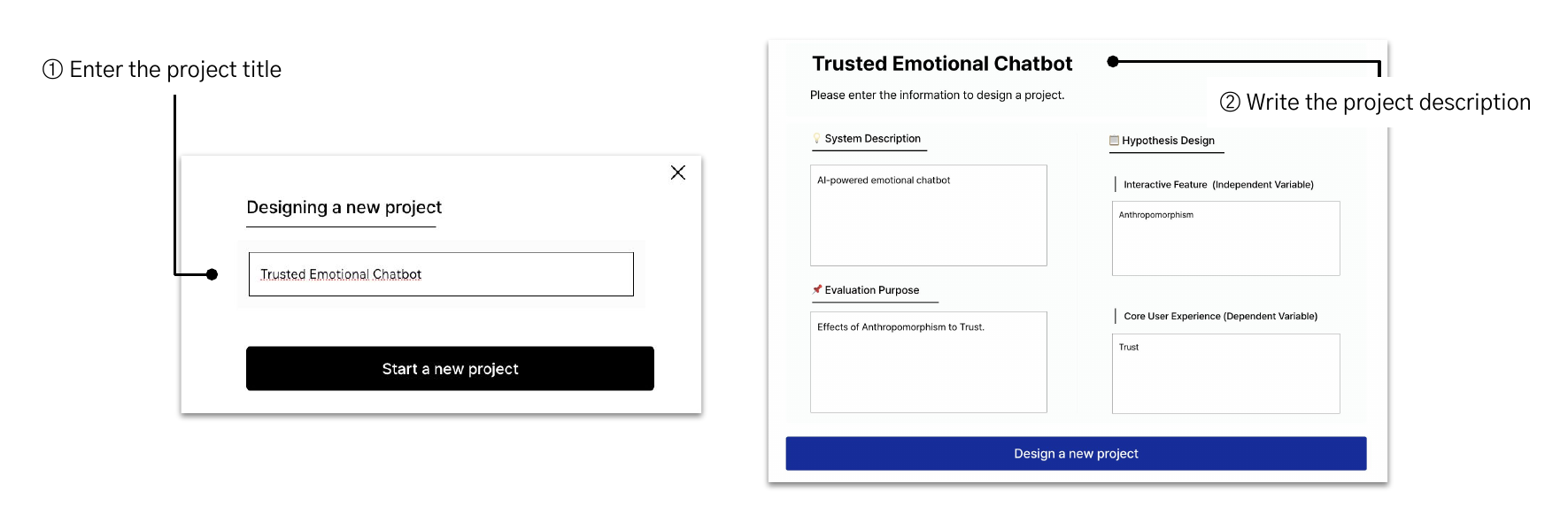}
  \caption{Project design page of UX Remix. In this page, researchers write a short project description.}
  \label{fig:project_design}
\end{figure}

On the project design page, the researchers fill in the basic project description about their project (see Figure~\ref{fig:project_design}). They first enter the project title. Then, they write the description of the system, the purpose of the evaluation, and the hypothesis design (the interactive features of the system and the core user experience). In the system description section, they enter the one-line description of the system or product being evaluated (e.g., AI-powered emotional chatbot). The evaluation purpose section receives the key aspects they want to explore through research (e.g., \textit{``I want to study how the anthropomorphism of an AI chatbot affects users' trust''}).

After that, researchers can fill in the hypothesis design section based on the evaluation purpose. The hypothesis design section includes interactive features and core user experience, which correspond to independent and dependent variables in a hypothesis, respectively. For instance, if the evaluation purpose is \textit{``I want to study how the anthropomorphism of an AI chatbot affects users' trust''}, then the interactive feature could be \textit{``anthropomorphism''}, and the core user experience could be \textit{``trust''}.

Once all sections are entered, the system stores the basic project description, which will be used to recommend the appropriate constructs and measurement items.

\subsubsection{Construct Selection Page}

% On the Metric Selection Page, researchers와 Practitioners는 시스템이 추천하는 Metric을 기반으로 현재 프로젝트의 UX evaluation purpose에 어울리는 metric을 선택하게 됩니다. page의 상단에는 시스템이 Project Design Page에서 입력한 detail 정보를 보여줍니다. 하단에는 입력한 interactive feature 정보와 UX evaluation construct 정보를 기반으로 Evaluation Goal이 생성되고, 그 아래에는 이에 대한 recommended metric이 보여집니다.

% recommended metric은 사용자가 입력한 evaluation purpose와 UX evaluation construct 정보를 토대로, 이전에 수집한 697개의 metric에서 가장 연관성이 높은 10개의 metric을 추천하여 보여줍니다. 이를 위해서 이전에 metric 정보가 저장된 vector database로부터 evaluation purpose와 UX evalaution construct와 가장 cosine similarity가 높은 metric을 검색합니다. 먼저, UX evaluation construct와 cosine similarity가 높은 20개의 metric을 검색합니다. 이어서 20개의 metric 중 evalaution purpose와 cosine similarity가 높은 10개의 metric을 검색해서 반환합니다. vector database가 반환한 metric은 the Metric Selection Page의 하단에 table 형태로 보여집니다. Table에서 각 metric name, metric definition, metric usage를 볼 수 있고, 마지막 행에 이 metric을 사용할 지 여부를 check할 수 있습니다. 각 metric의 행을 클릭하면, modal window가 overlay하여 각 metric에 대한 detail을 볼 수 있습니다. modal window에는 metric name, metric definition, metric usage 뿐만 아니라, metric의 point와 scale과 measurement items이 포함됩니다.

% researchers와 practitioners는 system이 추천한 metrics과 그 details을 토대로 자신의 research에 사용할 metrics을 선택합니다. 선택된 metrics은 Item Construction Page에서 custom metric과 definition을 설계하고, measurement items을 설계하는데 이용됩니다.

\begin{figure}[h]
  \centering
  \includegraphics[width=0.95\linewidth]{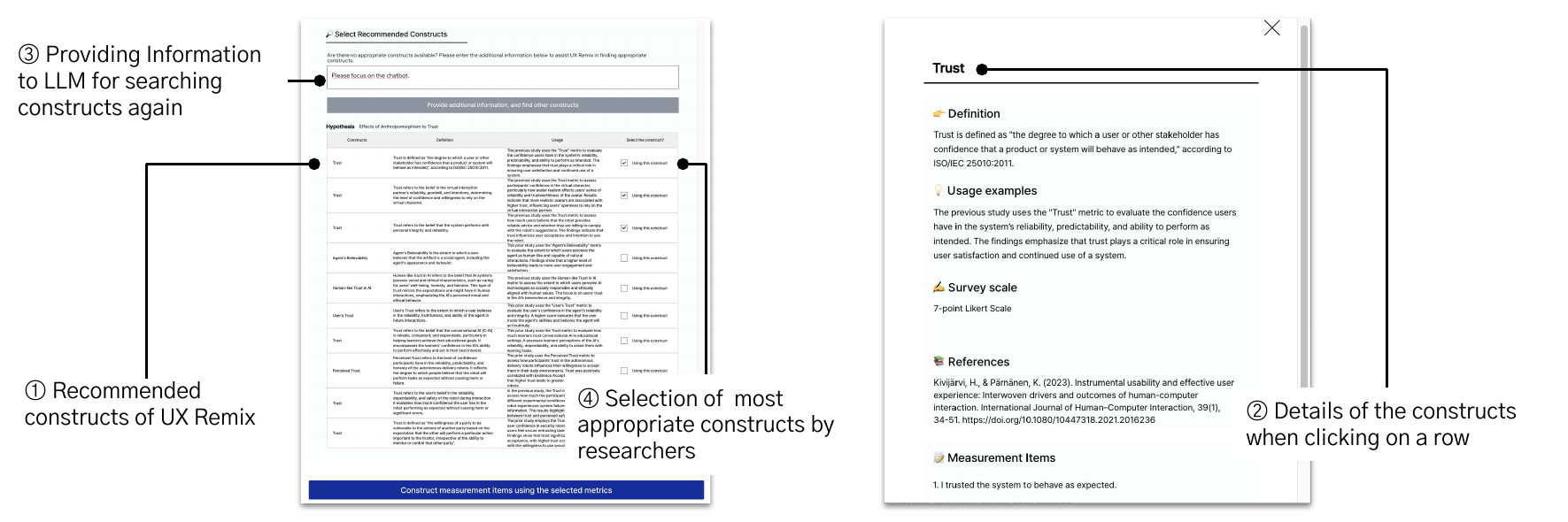}
  \caption{Construct selection page of UX Remix. UX Remix recommends 10 constructs relevant to the project description. Then, researchers select appropriate constructs based on the details.}
  \label{fig:construct_selection}
\end{figure}

On the construct selection page, researchers select the most appropriate constructs for their evaluation purposes based on the recommendations of the system (see Figure~\ref{fig:construct_selection}). First, the hypothesis is generated using the interactive feature provided and the core user experience. The recommended constructs are then displayed under hypothesis.

% 추천 및 선정 프로세스 더 자세히 설명하기
The 10 most relevant constructs are recommended based on the evaluation purpose and the core user experience. To achieve this, the system queries a vector database, retrieving the constructs with the highest cosine similarity to the evaluation purpose and the core user experience. Initially, the system searches for 20 constructs with the highest cosine similarity to the core user experience. After that, among these, 10 constructs with the highest cosine similarity to the evaluation purpose are selected and presented.

The constructs retrieved from the vector database are displayed in a tabular format at the bottom of the construct selection page. The table includes columns for the construct name, the construct definition, and the usage of the construct, with a final column allowing researchers to select whether to include the construct in their evaluation. Clicking on a row in the table shows a modal window overlay that provides detailed information about the clicked construct, such as measurement items, points, and scale type. %The modal window includes the construct name, the construct definition and the construct usage, as well as details about the measurement items and their points and scale type.

% 만약 제시된 Constructs에서 appropriate한 constructs가 없는 경우, researchers는 system에게 additional information을 제공하여, 이 정보를 포함하여 constructs를 검색할 수 있다. 사용자가 체크한 construct는 유지되며, 그 외의 constructs는 새로운 retrieval results로 교체된다. 이를 통해서 researchers는 자신에게 적합한 construct를 다시 추천받을 수 있다. 

If there is no appropriate constructs, researchers can provide additional information to the system, which is then utilized to retrieve new relevant constructs. The constructs already selected by the researchers are retained, while the other constructs are replaced by newly retrieved constructs. This approach enables researchers to receive recommendations for constructs that are more suitable to their research contexts. 

The selected constructs are used in the item development to customize a construct and design measurement items.

\subsubsection{Item Development Page}

\begin{figure}[h]
  \centering
  \includegraphics[width=0.95\linewidth]{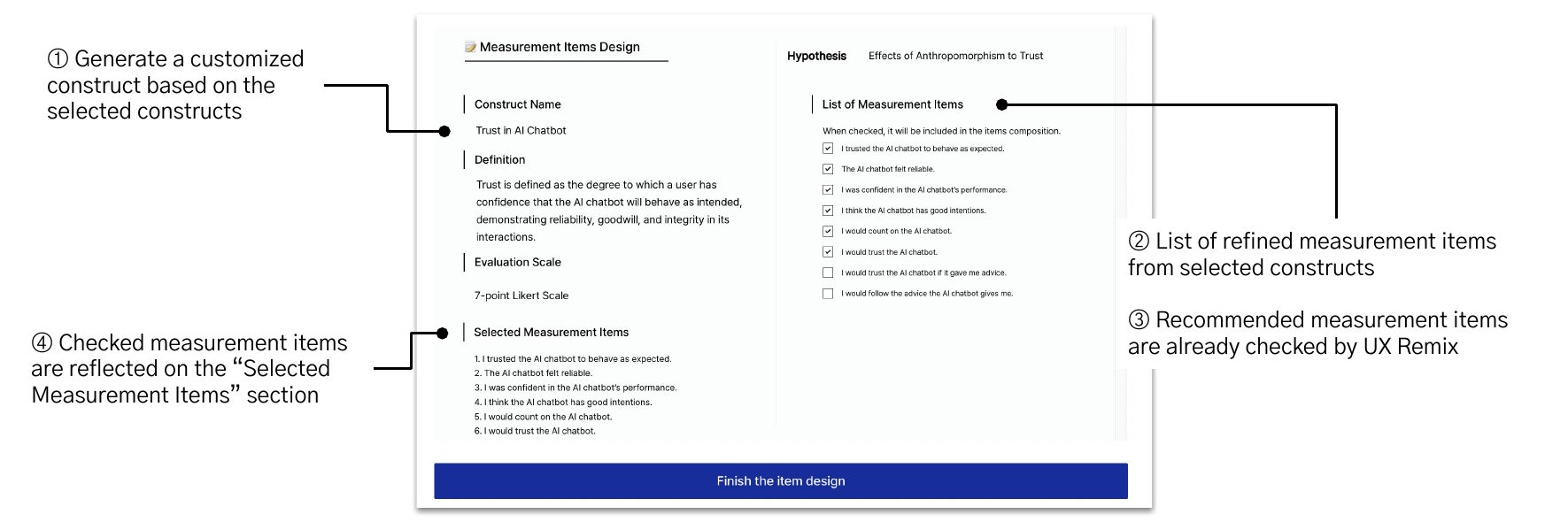}
  \caption{Item development page of UX Remix. UX Remix first generates a custom construct based on the selected constructs from previous page. Then, it refines the measurement items from the selected constructs and recommends most appropriate items to the custom construct. After that, researchers could decide most appropriate items based on the recommendation results.}
  \label{fig:item_development}
\end{figure}

On the item development page, researchers design a custom construct and corresponding measurement items tailored to the purpose of the evaluation based on the selected constructs (see Figure~\ref{fig:item_development}). At the top of the page, project details and selected constructs are displayed. At the bottom, the name, definition, point, and scale of the custom construct and measurement items are generated based on the selected constructs. Moreover, measurement items are modified to fit the current research context and presented as a checkable list.

% As mentioned above, the system generates a custom construct, modifies measurement items to align with the current research context, and recommends measurement items appropriate for the evaluation purpose. To achieve this, 
The system utilizes an LLM on the back-end to generate the custom construct and modify the measurement items. We have employed a Google Gemini 2.0 Flash model as an LLM. The name, definition, point, and scale of the custom construct are generated by the LLM based on the names, definitions, points, and scales of selected constructs (see Appendix~\ref{appendix:gen_construct} for detailed prompts). For measurement items, the LLM considers system description, evaluation purpose, interactive features, core user experience, hypothesis, custom construct name, and custom construct definition to refine the items for the current research context (See Appendix~\ref{appendix:item_refine} for detailed prompts). In this process, the meaning and expression of the items are preserved, while irrelevant or unnecessary elements are modified or removed. Following refinement, the LLM distinguishes the appropriate and inappropriate items for current research purposes (see the Appendix~\ref{appendix:item_recommend} for detailed prompts). The measurement items classified as appropriate are recommended by the system. The recommended items are pre-checked in the list of measurement items.

Researchers can review the custom construct and verify the modified measurement items. Furthermore, considering the recommended measurement items, they can further select the most suitable items or remove the unsuitable items for their research contexts and purposes. Through the procedure, users can design custom constructs and measurement items that are appropriate to their research contexts based on previous studies.

\section{Discussion and Limitations}

\subsection{Tailoring Custom Measurement Items for Evaluation using LLMs}
% LLM을 활용하여 연구 맥락에 맞는 Metric 설계 및 Measurement Item 설계 가능성

% Recent works on UX research design with LLM, LLM은 현재 연구 맥락에 맞는 metric을 추천할 수 있었다. 더 나아가, 이 논문에서는 주어진 정보와 이전의 연구 내용을 토대로, LLM이 다양한 metric와 measurement items을 기반으로 현재 연구 맥락에 맞는 cutomized metric과 measurement items을 설계할 수 있다는 것을 확인하였다. 이는 HCI 및 UX 리서치 및 평가에서 어렵고 명확한 프로세스가 존재하지 않았던 metric과 measurement items 설계 과정을 이전의 연구 문헌과 LLM을 활용하여 개선할 수 있는 가능성을 보여준다. In particular, measurement items의 경우 이전 연구에서 사용한 measurement items을 현재 연구 맥락에 맞게 수정해야 하는 경우가 많은데, LLM을 활용하면 이러한 과정에서 필요한 measurement item 선택과 구성, 수정 과정을 단축시킬 수 있다. 뿐만 아니라, usability, usefulness, ease-of-use와 달리 많이 사용되지 않는 UX construct의 측정을 위한 metric에 대한 정보는 찾기가 어려운데, vector database 기반의 search를 통해서 비슷한 metric을 빠르게 탐색하고, 이러한 metric을 활용하여 measurement items을 구성할 수 있어 새로운 UX metric과 Measurement Item의 설계에 도움을 줄 수 있다.

Recent work demonstrated that LLM can help explore metrics that align with the research context \cite{zheng2024evalignux}. Furthermore, in this paper, we confirm that LLMs can design measurement items that fit to the current research context by leveraging a diverse set of constructs and measurement items. This finding suggests the potential to improve the measurement item design process in HCI research, which traditionally lacks a clear and standardized process \cite{perrig2024measurement, law2014attitudes, hornbaek2006current, lindgaard2013introduction}, using the prior literature and LLM.

In particular, adaptation of measurement items from previous studies is often required to fit the current research context \cite{muller2014survey}. The use of LLMs can shorten this process by supporting the selection, configuration, and modification of measurement items. Moreover, unlike commonly used constructs, such as usability, usefulness, and ease-of-use, it is challenging to find measurement items for less frequently measured constructs. Through leveraging vector database-based search, researchers can rapidly find similar constructs and utilize them to develop measurement items, thereby aiding the design of constructs and measurement items.

\subsection{Enhancing Reliability and Rigor of Measurement Item Design}
% 이전 연구 Metric을 재사용하거나 혼합하여 사용하여 self-development보다 엄밀성과 신뢰도 향상 가능성

% 이전의 연구에서 많은 연구에서 사용된 Metric이 거의 재사용되지 않거나, 비슷한 metric이 있더라도 self-development metric을 사용하는 경우가 많았다. 이는 이전 연구와 맥락이 다르거나, 재사용할만한 Metric을 찾지 못했기 때문이다. 이러한 이유로 많은 연구에서 사용된 검증된 metric과 measurement items이 다른 연구에서 사용되지 못하고 있으며, 전통적인 UX metric에 의존하는 경우가 많다. 특히 Self-Development된 metric이나 measurement item의 경우에는 이전 연구에 비해 맥락에 맞출 수 있는 자율성은 올라가지만, 상대적으로 과학적 엄밀성이나 신뢰도가 떨어질 수 있다. 따라서 기존에 사용된 metric을 재사용하거나 이를 적절히 혼합하여 새로운 metric을 커스텀하는 것은 self-development item보다 상대적으로 과학적 엄밀성이나 신뢰도를 높일 가능성이 있다. 우리의 UX Remix는 연구자나 pratitioners가 이전 연구의 metric과 measurement items을 기반으로 자신의 연구를 위한 metric과 measurement items을 설계하도록 지원한다. 이를 통해, evaluation 과정에서 놓치기 쉬운 metric과 measurement items의 엄밀성과 신뢰도를 상대적으로 더 보장할 수 있을 것이라 기대한다.

According to previous literature, measurement items have rarely been reused, and even when similar constructs exist, researchers often develop their own measurement items \cite{perrig2024measurement, bargas2011old}. The reasons for the tendency are due to differences in research contexts or the difficulty in identifying appropriate reusable measurement items \cite{perrig2024measurement}. Therefore, the pre-defined measurement items are underutilized in other studies, and the research often rely on traditional questionnaires. In particular, self-developed measurement items may have greater flexibility to contextualize, but it might be less rigorous or reliable compared to established or pre-defined measurement items.

However, reusing existing measurement items or appropriately combining them has the potential to improve rigor and reliability compared to self-developed measurement items \cite{kimberlin2008validity, fisher2016developing}
. Our system aims to assist researchers in designing constructs and measurement items based on previous constructs and measurement items, enabling them to tailor measurement items to specific research needs. Through the system, we anticipate greater assurance of the rigor and reliability of measurement items that are often overlooked in the evaluation.

\subsection{Limitations and Future Work}

% 우리의 work는 limitations과 보완할 점이 있습니다. % 먼저, LLM을 활용하여 construct를 customize하고, meausrement items을 design하는 프로세스에서 기존 LLM의 학습 데이터에 영향을 받을 수 있습니다. 따라서 향후에는 이러한 기존 데이터에 대한 편향에 대해 탐구할 필요가 있습니다. 또한, 현재 custom construct를 생성하고, measurement items을 design하는 과정은 기존 construct를 적절히 병합하고, items의 의미를 헤치지 않으면서 context에 맞게 바꾸는 일부 rule을 기반으로 한 prompt에 의존하고 있습니다. 향후에는 construct와 measurement items의 rigor를 유지하기 위한 더 상세한 rule을 기반으로한 prompt design이 필요합니다. 
% 시스템에도 여러 limitations이 존재합니다. 현재 저희가 개발한 UX Remix 시스템은 실제 HCI 및 UX 리서처 등을 대상으로 User Study나 검증이 진행되지 않았습니다. 실제로 UX Metric 설계와 UX Evaluation에 도움을 주고 효과를 검증하기 위해서는 HCI 연구자와 pratitioners를 대상으로 시스템에 대한 피드백을 듣고 이를 바탕으로 개선해나가는 iterative process가 필요합니다. 우리의 시스템은 아직 충분한 수의 metric과 measurement items이 확보되지 않았습니다. 넓은 범위의 UX Evaluation을 cover하기 위해 여전히 다양한 평가지표를 확보할 필요가 있으며, 만약 적합한 평가지표가 없을 때는 어떻게 design process 지원할 수 있을지에 대한 고려가 필요합니다. 또한, 우리의 연구는 현재 survey기반의 metric 설계와 measurement items의 구성만을 지원하고 있습니다. 향후에는 LLM을 기반으로 measurement items의 신뢰도나 통계 분석 방법, 설계한 metric과 measurement items을 토대로 evaluation 및 study plan을 설계하는 방법 등을 지원하는 방법도 고려할 수 있습니다. 마지막으로 설계한 metric과 measurement items의 과학적 엄밀성과 신뢰도를 높이는 방법을 마련해야 합니다. 현재는 단순히 previous paper에 의존하여 신뢰도를 보장하고 있지만, 더 나은 방법이 필요합니다. 가령, crowdsourcing 방법을 활용하여 다른 사람들이 사용한 metric이나 measurement items을 재사용하거나, 이들이 설계에 대한 피드백, 실제 연구 데이터에 기반한 피드백으로 meausrement item의 추천 방법을 개선할 수 있습니다. 이렇듯 다양한 researchers와 practitioners의 feedback에 기반하여 custom metric과 measurement items의 과학적 엄밀성과 신뢰도를 높이는 방법을 고안해야 합니다.

Our work has several limitations and areas for improvement. First, the process of customizing constructs and designing measurement items using LLMs is inherently influenced by training data. Therefore, future research should consider potential biases and prevent them when using LLMs to design the measurement process. Additionally, the current approach to generating custom constructs and designing measurement items relies on prompts with specific rules that guide the appropriate merging of existing constructs and the contextual adaptation of measurement items without compromising their original meaning. It will be necessary to develop more refined prompt designs that reflect the survey validation process to ensure the rigor of both the constructs and the measurement items.

The system also has several limitations. Our system has not yet been evaluated and validated by user studies with researchers. To effectively support the questionnaire design process, we should collect feedback from them and refine the system accordingly. Moreover, our system currently lacks a sufficient number of constructs and measurement items. To cover a broader range of evaluations, it is essential to collect a diverse set of constructs and measurement items. Additionally, we should consider how to support the questionnaire design process when appropriate constructs and measurement items are not available. Furthermore, our system currently supports only construct selection and development of measurement items. In the future, it would be beneficial to explore methods to validate the reliability of measurement items and statistical analysis techniques, and ways to support evaluation planning using the custom constructs and measurement items. Finally, we should improve the rigor and reliability of the designed constructs and measurement items. Currently, we primarily rely on previous research to ensure reliability and rigor, but better methods are needed. For instance, crowdsourcing approaches could be used to improve the reliability of measurement items based on feedback. %Thus, it is crucial to develop methods that enhance the rigor and reliability of custom metrics and measurement items based on feedback from researchers.
% \section{Use Cases}

\section{Conclusion}

In this paper, we present UX Remix, a system that supports the measurement item design process using LLMs. We first collected 697 constructs from the previous literature and extracted relevant information using LLMs. Then, we pre-processed the measurement items for each construct and stored them in a vector database. UX Remix uses data to support the design process for custom constructs and measurement items. UX Remix first receives a project description that includes details on the system description, the purpose of the evaluation, and the hypothesis. Then, UX Remix retrieves the 10 most relevant constructs from the vector database based on the project description and recommends them. Afterward, UX Remix refines the construct and measurement items according to the project description and the selected constructs. Finally, UX Remix suggests appropriate measurement items aligned with the project context. Through the process, we expect that researchers could design custom constructs and measurement items based on pre-defined measurement items that fit to their research contexts. 

\bibliography{references}

\appendix


% The sources for the ceur-art style are available via
% \begin{itemize}
% \item \href{https://github.com/yamadharma/ceurart}{GitHub},
% % \item \href{https://www.overleaf.com/project/5e76702c4acae70001d3bc87}{Overleaf},
% \item
%   \href{https://www.overleaf.com/latex/templates/template-for-submissions-to-ceur-workshop-proceedings-ceur-ws-dot-org/pkfscdkgkhcq}{Overleaf
%     template}.
% \end{itemize}

\section{Prompts}

\subsection{Design GPT for Data Extraction}\label{appendix:gpt_design}

Below is the description and instructions of the GPT to extract construct and information from collected papers.

\noindent\hrulefill
\newline
\textbf{Description}
\begin{itemize}
    \item Based on the given paper and construct, the GPT returns the construct definition, construct usage, point and type of measurement items, measurement items, number of items, title of the paper, and an APA reference
\end{itemize}

\noindent\textbf{Instructions}
\begin{itemize}
    \item This GPT provides the construct definition, construct usage, point and type of measurement items, measurement items, number of items, title of the paper, and an APA reference based on the given paper and construct. The output should follow the format below.
    \begin{itemize}
        \item \textbf{Construct:} [Construct Name]
        \item \textbf{Definition:} [Definition of the construct, \textit{Construct Name} is ...]
        \item \textbf{Usage:} The paper uses the [X] to evaluate users' [Y] regarding [Z technology]. The paper found [findings related to Y]. \cite{zheng2024evalignux}
        \item \textbf{Point and Type of Measurement Items:} [7]-point Likert Type, [6]-point Semantic Type, ...
        \item \textbf{Measurement Items:} [1. ... | 2. ... |  ...]
        \item \textbf{Number of Measurement Items:} [The number of measurement items specified in the paper.]
        \item \textbf{Title:} [Title of the paper]
        \item \textbf{APA reference:} [APA format of the paper]
    \end{itemize}
\end{itemize}
\noindent\hrule

\subsection{Prompts for Custom Construct Generation}\label{appendix:gen_construct}

Below is a prompt for generating a custom construct. The texts inside curly braces (i.e. `` \{ \} '') indicate variable names. 

\noindent\hrulefill
\newline
\textbf{Role: system} \\
You are an expert in HCI research methodology. Please customize the construct name, definition, point, and type based on the given construct names, definitions, points, and types. Please consider the given system description, evaluation purpose, and hypothesis. You must respond using the defined JSON schema.\\

\noindent\textbf{Role: user} \\
System description: \{System description\}\\
Evaluation purpose: \{Evaluation purpose\}\\
Hypothesis: Effects of \{Interactive feature\} to \{core user experience\} \\
     
\noindent Selected constructs name: \{List of selected constructs' name\}\\
Selected constructs definition: \{List of selected constructs' definition\}\\
Selected constructs point (scale): \{List of selected constructs' point\}\\
Selected constructs type (scale type): \{List of selected constructs' scale type\}\\
        
\noindent  Please customize the construct name, definition, point, and type based on the given construct names, definitions, points, and types.\\
\noindent Please consider the given system description, evaluation purpose, and hypothesis.\\
Response format: \\
1. Custom construct name: (string of custom construct name) \\
2. Custom construct definition: (string of custom construct definition) \\
3. Custom construct point (scale): (integer of custom construct point) \\
4. Custom construct type (scale type): (string of custom construct type) \\
\noindent\hrule

\subsection{Prompts for Refinement of Measurement Items}\label{appendix:item_refine}

Below is a prompt for refinement of measurement items. The texts inside curly braces (i.e. `` \{ \} '') indicate variable names. 

\noindent\hrulefill
\newline
\textbf{Role: system} \\
You are an expert in HCI research methodology. Please refine the measurement items to make them based on given context. You must respond using the defined JSON schema.\\

\noindent\textbf{Role: user} \\
System description: \{System description\}\\
Evaluation purpose: \{Evaluation purpose\}\\
Interactive feature: \{Interactive feature\}\\
Core user experience: \{core user experience\} \\
Hypothesis: Effects of \{Interactive feature\} to \{Core user experience\} \\
Construct Name: \{Custom construct name\} \\
Construct Definition: \{Custom construct definition\}\\

\noindent Available measurement items: \{Measurement items from selected constructs\}\\
     
\noindent Based on the context, please modify the measurement items to be more natural and clear. Please follow below rules.\\
1. Please change the terms like [Evaluation Target] to appropriate words in all measurement items. After that, if the measurement item sentence is awkward, please modify it to be appropriate.\\
2. As you know, the construct name and core user experience might not be evaluation target. \\
3. If you think that the measurement item is a reverse-coded item, Please add (R, reverse) after the reverse-coded item.\\
4. Please exclude any words in the measurement items that do not fit (e.g. term like ``educational content'' could be excluded if the evaluation purpose or product explanation is not for education).\\
5. However, please maintain the original meaning and expression as much as possible, and only remove unnecessary or inappropriate parts. \\
6. Furthermore, please maintain the format of the measurement items.\\

\noindent For instance, ``I am satisfied with visiting this website.'' to ``I am satisfied with using this app.'', ``I believe in the educational capabilities of avatar.'' to ``I believe in the capabilities of avatar.''\\

\noindent\hrule

\subsection{Prompts for Classification of Measurement Items}\label{appendix:item_recommend}

Below is a prompt for classification of measurement items. The texts inside curly braces (i.e. `` \{ \} '') indicate variable names. 

\noindent\hrulefill
\newline
\textbf{Role: system} \\
You are an expert in HCI research methodology. Please select the most appropriate measurement items and the inappropriate measurement items based on the evaluation purpose and hypothesis. You must respond using the defined JSON schema.\\

\noindent\textbf{Role: user} \\
System description: \{System description\}\\
Evaluation purpose: \{Evaluation purpose\}\\
Interactive feature: \{Interactive feature\}\\
Core user experience: \{Core user experience\} \\
Hypothesis: Effects of \{Interactive feature\} to \{Core user experience\} \\
Construct Name: \{Custom construct name\} \\
Construct Definition: \{Custom construct definition\}\\

\noindent Available measurement items: \{Refined measurement items\}\\
     
\noindent Please select the most appropriate measurement items and the inappropriate measurement items based on the evaluation purpose and hypothesis.\\
Please maintain the format of the measurement items.\\
Please respond in sentences rather than numbers.\\
Response format:\\
1. Appropriate items: (list all appropriate items in one line, separated by |)\\
2. Inappropriate items: (list all inappropriate items in one line, separated by |)\\
3. Rationale for selection: (string of rationale for selection)\\

\noindent\hrule

\end{document}